# Diffusion processes in germanium and silicon films grown on Si₃N₄ substrates

Larisa V. Arapkina[1], Kirill V. Chizh[1], Dmitry B. Stavrovskii[1],

Vladimir P. Dubkov[1], Elizabeth P. Lazareva[2], and Vladimir A. Yuryev[1]

[1]Prokhorov General Physics Institute of the Russian Academy of Sciences, 38 Vavilov Street, 119991 Moscow, Russia

[2]Institute of Nanotechnology of Microelectronics of the Russian Academy of Sciences, 32A Leninsky Prospekt, 119991 Moscow, Russia

Corresponding author:

Larisa V. Arapkina, e-mail: arapkina@kapella.gpi.ru

## Abstract

In this article, the results of investigation of processes occurring during the molecular-beam deposition of germanium layers on $Si_3N_4$ dielectric substrates within a wide range of the Ge film growth temperatures (30 to 600°C) are presented. The intensity of the IR absorption bands related to the vibrations of the N–H and Si–N bonds are established to decrease with the increase of the Ge deposition temperature. This phenomenon apparently cannot be explained as only a thermally activated process. Simultaneously, the peaks corresponding to the Ge–N and Ge–O vibration bonds emerge in the X-ray photoelectron spectra. We believe that the deposition of Ge films on $Si_3N_4$ dielectric substrates containing hydrogen atoms resulted in the migration of hydrogen atoms from the dielectric layer into the growing film. We interpret the experimental results in terms of a model, according to which the diffusion of hydrogen atoms from the $Si_3N_4$ layer into the growing Ge film occurs due to the difference in chemical potentials of hydrogen atoms in the dielectric layer and the germanium film. This process initiates the diffusion of germanium atoms in the opposite direction, into the $Si_3N_4$ layer, where they connect to dangling bonds of nitrogen atoms arising due to the escape of hydrogen atoms. The analogous processes occur during the deposition of silicon layers on $Si_3N_4$ substrates.

**Key words**

Silicon

Germanium

Hydrogen diffusion

Semiconductors (IV) on dielectric

Optical properties

Spectroscopy

## 1. Introduction

In a state-of-the-art technology, silicon is the main material for manufacturing semiconductor structures, however, for some applications, germanium may be more suitable. Crystalline germanium, due to its narrow band gap, is promising to be used in high-efficiency solar panels based on multiple junctions [1] for providing the efficient conversion of solar energy. The structures presented in Ref. [1] demonstrated good exploitation characteristics; however, expensive substrates of crystalline germanium are required for their fabrication. In Refs. [2],[3], solar cell fabricated by deposition of a thin low resistivity polycrystalline germanium film (poly-Ge) on glasses covered with a $Si_3N_4$ layer have been demonstrated. The advantage of this approach is that there is no necessity to use a transparent conductive oxide layer as a contact in these structures. Additionally, those structures are resistant to ultraviolet radiation for a long time.

Another important application of poly-Ge is its employment in three-dimensional sequential-layered integration structures as the underlying interconnects. Since the reduction of the thermal budget of the channel preparation and the source-drain activation are known to be critical issues in three-dimensional sequential-layered integration structures [4], the formation of poly-Ge layer on $Si_3N_4$ that requires a lower deposition temperature and exhibits higher carrier mobility than poly-Si appears to be preferable compared to the deposition of poly-Si [4]. In Ref. [5], the authors have proposed using the MEMS structural layer consisting of poly-Ge directly as the interconnecting material between MEMS and an electronic circuit. The authors claim that poly-Ge due to the higher conductivity is more appropriate for the majority of applications in the MEMS technology than amorphous silicon (α-Si) [5]. It is worth mentioning the application poly-Ge in hybrid plasmonic-photonic nano-ribbon waveguides. The authors of Ref. [6] have supposed using a poly-Ge infrared waveguide based on the channel completely embedded in $Si_3N_4$ that is expected to be compatible with CMOS processing.

All the above-mentioned technical solutions utilize poly-Ge formed on a $Si_3N_4$ film; therefore, it is highly important to study the processes that occur when Ge is deposited on a $Si_3N_4$ layer. In this article, we study the processes happening during the deposition of germanium layers on dielectric $Si_3N_4$ substrates in a wide range of growth temperatures varied from 30 to 600°C. In our previous article [7], we have studied the deposition of the Si films on the $Si_3N_4$ substrates and observed the hydrogen atoms diffusion from the dielectric layers into the growing films at low temperatures. We have supposed that similar processes could also occur in the case of the Ge layer growth. In the current work, we have explored Ge films deposited on $Si_3N_4$ substrates by means of Fourier transform infrared (FTIR), Raman and X-ray photoelectron (XPS) spectroscopy. It was found that not only the diffusion of hydrogen atoms from $Si_3N_4$ substrates takes place but also germanium atoms diffuse from the growing film into the dielectric layer. The results obtained allowed us to propose a model explaining the observed phenomenon.

## 2. Sample preparation, experimental methods and equipment

### 2.1. Samples

The samples were germanium layers coating $Si_3N_4/SiO_2/Si(100)$ dielectric substrates. The Ge layers were deposited from molecular beam (MB) on $Si_3N_4/SiO_2/Si(100)$ dielectric substrates in a Riber EVA32 ultra-high vacuum (UHV) molecular beam epitaxy (MBE) chamber. The multilayer dielectric substrates were prepared as described in our previous articles [7]–[11]: 530-nm thick $SO_2$ layers were obtained by thermal oxidation of $Si(001)$ substrates, whereas 175-nm thick $Si_3N_4$ layers were formed by chemical vapor deposition (CVD). It should be noticed that since the upper $Si_3N_4$ layer was deposited by CVD using pyrolysis of a monosilane-ammonia mixture at the temperature of 750°C for 60 min [7]–[11] it can contain up to 8% of H atoms [12],[13]. Before depositing the Ge layers, the dielectric substrates were treated in the ammonia-peroxide solution ($NH_4OH$ (27%): $H_2O_2$ (30%): $H_2O$ [1:1:3], boiling for 10 min), then rinsed in deionized water, boiled in high-purity isopropyl alcohol ([$C_3H_7OH$] > 99.8 wt.%, $T \approx 70$ °C), and dried in the isopropyl alcohol vapor (for 10 min) and the clean air. Additionally, before moving into the MBE chamber, the substrates were annealed at 600°C at the residual gas pressure of less than $5\times10^{-9}$ Torr in the preliminary annealing chamber for 6 hours. It is worth mentioning right away that FTIR spectra of the substrate recorded before and after each step of the preliminary treatment (chemical treatment in the ammonia-peroxide solution and the preliminary annealing at 600°C) were completely identical to each other. The MBE chamber was evacuated down to about $3\times10^{-11}$ Torr before the processes. The pressure did not exceed $5\times10^{-9}$ Torr during the Ge deposition. The Ge layers were deposited from electron beam source. The deposition rate was measured using Inficon XTC751-001-G1 (Leybold-Heraeus) film thickness monitor. The Ge deposition rate was ~ 0.25 Å/s. The film growth temperature was varied from 30 to 600°C. The thickness of the Ge layers was 200, 50 nm and 20 Å. The samples with 50-nm

Ge layers were annealed at 500°C after the flux shutoff of Ge atoms, so as the total heat treatment times of 200 nm and 50 nm thick films to be equal. The growth cycle of 20 Å thick Ge layers at 500°C comprised two steps—growing and annealing—and the total duration of the process was 15 min.

During the production, samples were heated from the rear side using a tantalum spiral heater. The temperature was measured with thermocouples of the heaters, which were in situ graduated beforehand against the IMPAC IS12-Si infrared pyrometer (LumaSense Technologies). The residual atmosphere composition in the MBE chamber was monitored using the RGA-200 residual gas analyzer (Stanford Research Systems) before and during the deposition process. The crystal structure of the growing layers was *in-situ* studied directly during the deposition using reflected high-energy electron diffraction (RHEED) tool (Staib Instruments).

## 2.2. FTIR analysis

Fourier-transform IR (FTIR) transmission and reflection spectra were registered using a vacuum IFS-66v/S spectrometer (Bruker). The spectral resolution was 10 cm$^{-1}$ that enabled both an adequate recording of all spectral features and filtering away the high-frequency components related to light interference in silicon wafers. An opaque golden mirror was used as a reference sample during recording of reflectance spectra. The instrument was evacuated to the residual air pressure of 2 mbar during spectra recording that enabled a considerable reduction of spectral interferences associated with carbon dioxide and water vapor. Direct analysis of the obtained spectra was hampered with the interference of the probe radiation in thin layers of materials on silicon wafers. The procedures of IR spectra exploring and processing were similar to those used by us in Ref. [7].

## 2.3. Raman spectroscopy

Raman spectra were obtained with a Renishaw inVia Qontor spectrometer. For spectra acquisition, vertically polarized laser excitation at the wavelengths of 632.8 and 785 nm, a diffraction grating with 1200 grooves/mm and a 50× objective were employed. The laser emission power was 1.7 or 5.4 mW for 632.8 and 785-nm lasers, respectively. The duration of each measurement was 10 s, during which 20 or 30 spectra for averaging were obtained with 632.8 and 785-nm lasers, respectively. The measurements were carried out at room temperature. Gaussian or Lorentz functions were used for the peak analysis of Raman bands.

## 2.4. X-ray photoelectron spectroscopy

All the samples were also examined by means of the X-ray photoelectron spectroscopy (XPS). The experiments were carried out in a UHV analytical chamber of an SSC2 surface science center (Riber) equipped with a cylindrical mirror type (CMA) EA 150 electron energy analyzer (Riber) and an X-ray photoexcitation source. In the SSC2 instrument, the analytical chamber is connected with the EVA32 UHV MBE chamber (Riber) by a high-vacuum transfer line thus the samples never leave the high vacuum conditions during the entire production and analysis time that enables transposing the grown samples for the analysis practically without an exposure to the ambient.

The photoelectron spectra were recorded with the resolution of 1.8 eV for the survey scans and 0.62 eV for the high-resolution spectra. A non-monochromatic Al $K_{\alpha1,2}$ X-ray ($h\nu = 1486.6$ eV) source with the radiation power of 180 W was used to generate photoelectrons. An Ar$^+$-ion gun that ran at the ion beam energy $E \sim 3.5$ kV and the current $I \sim 0.08$ μA was employed for depth profiling.

The N1$s$, Si2$p$, Ge2$p$ and Ge3$d$ peaks were decomposed into a few components. The deconvolution procedure was held with the XPSPEAK 4.1 program, which approximated the photoelectron peaks with a product of Gaussian and Lorentz functions. Doublet structure of the non-monochromatic radiation was taken into account. Energy shifts of the peak positions related to elements in various chemical compounds were compared with the NIST X-ray Photoelectron Spectroscopy Database [14],[15] and the data of the original studies. The spectrometer was calibrated against the Si–N signal (binding energy $E_b = 398.6$ eV) and Ar signal ($E_b = 243$ eV) since the C1$s$ peak was too weak to use.

## 3. Results

### 3.1. Raman spectroscopy and RHEED

The Raman spectra of the samples with 200 nm thick Ge layers are depicted in Figure 1; for clarity, the spectra have been shifted vertically. The structure of Ge layers is seen to evolve with the growth temperature. The layers grew amorphous ($\alpha$-Ge) in the temperature range from 30 to 200°C and polycrystalline (poly-Ge) at $T > 300$°C that was confirmed by RHEED measurements carried out during the film growth.

For all amorphous layers, the bands assigned to the TA(Ge) ~ 80 cm$^{-1}$, LA(Ge) ~ 170cm$^{-1}$, LO(Ge) ~ 230 cm$^{-1}$, TO($\alpha$-Ge) ~ 270 cm$^{-1}$ vibrations [16],[17] have been observed. Transition to the polycrystalline structure has led to the reduction of the TA(Ge) vibration band intensity and narrowing of the LA(Ge) and LO(Ge) bands. The TO($\alpha$-Ge) vibration band has blueshifted and taken the position of the TO(c-Ge) vibration band (~ 299 cm$^{-1}$) observed in crystalline Ge. In addition, new bands peaked at ~ 352 cm$^{-1}$ (TO+TA), ~ 469cm$^{-1}$ (LO+LA) and ~ 569 cm$^{-1}$ (2TO) [18] have appeared in the spectra obtained at the polycrystalline Ge layers.

The amorphous and polycrystalline 200-nm thick Ge layers entirely absorbed the radiation of a 632.8-nm laser since the value of its absorption coefficient is large in this spectral range [19],[20]. The use of a 785-nm laser for the Raman scattering excitation in the poly-Ge layers led to the appearance of an additional weak band peaked at ~ 520 cm$^{-1}$ and assigned to the Si–Si bond vibration (TO c-Si) in the spectra, which was due to the penetration of the laser radiation through the 200-nm thick poly-Ge film and all the substrate dielectric layers followed by its Raman scattering within the Si substrate [19],[20]. The comparison of the Raman spectra of the initial Si$_3$N$_4$/SiO$_2$/Si(001) substrate and c-Si has shown that peaks related to the light scattering on Si–N and Si–O bonds had not been registered due to the presence of strong absorption bands that corresponded to the vibration of the Si–Si bonds.

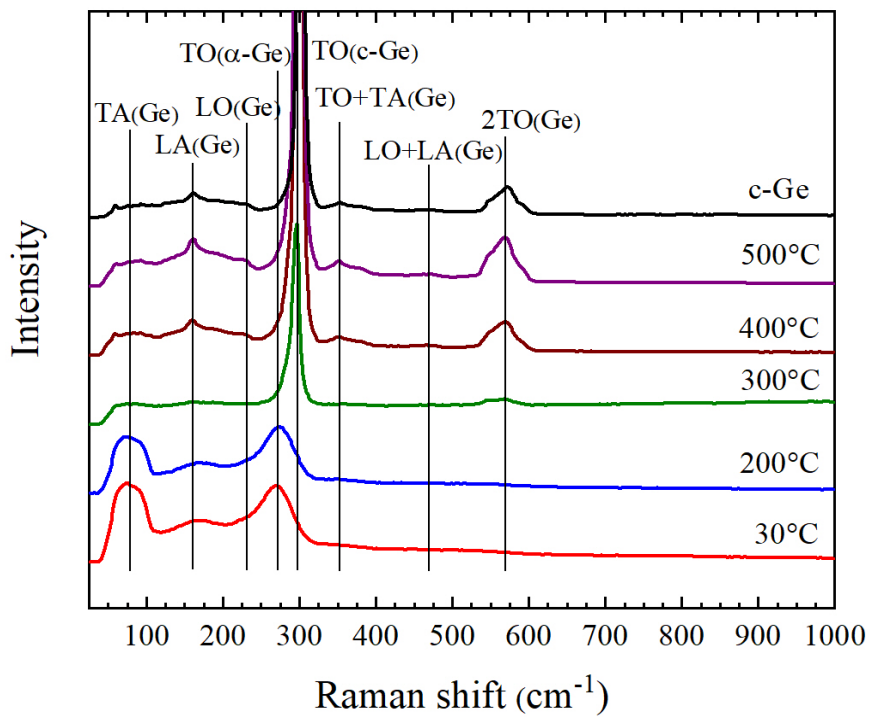

Figure 1. Raman spectra of Ge/Si$_3$N$_4$/SiO$_2$/Si(001) samples grown at different temperatures; the laser wavelength is 632.8 nm; the Ge layer thickness is 200 nm; the spectrum marked as c-Ge was recorded at a sample of crystalline Ge.

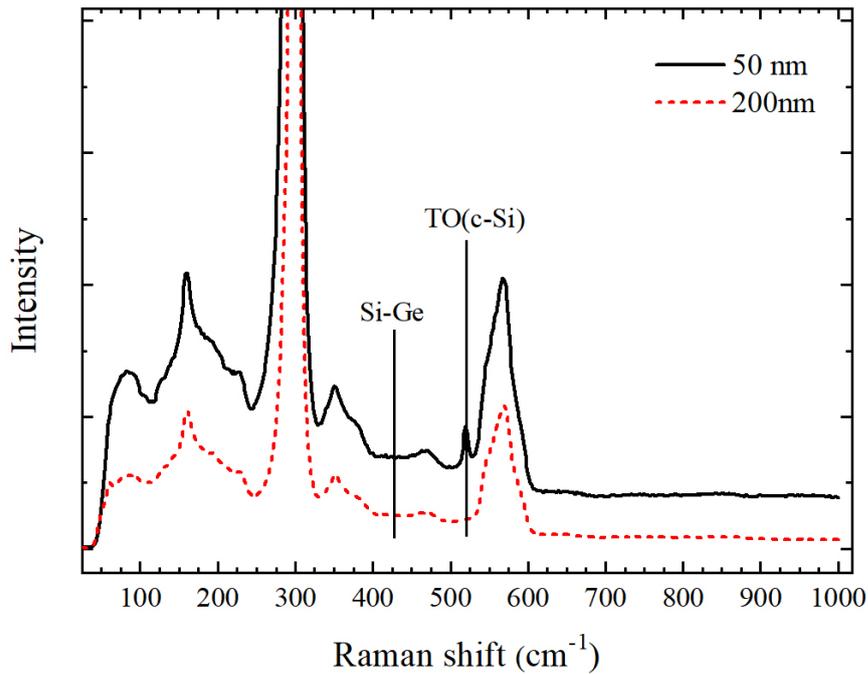

Figure 2. Raman spectra obtained at 200 and 50 nm thick poly-Ge layers grown at 500°C; the laser wavelength is 632.8 nm; the vertical line marked as Si–Ge points out an expected position of the band corresponding to the Si–Ge bond vibration.

Figure 2 demonstrates Raman spectra for Ge/Si$_3$N$_4$/SiO$_2$/Si(001) samples with the thicknesses of 200 and 50 nm grown at 500°C. The TO(c-Si) vibration band at ~ 520 cm$^{-1}$ connected with the scattering at the Si–Si bonds within the Si substrate has appeared in the spectra of the thinner samples with the 50-nm thick poly-Ge film. It should be noticed that the characteristic vibration bands of the Si–Ge bonds, which could peak around 400 cm$^{-1}$ [21],[22], have not been observed in the spectra (an expected position of the Si–Ge band is pointed out by the left vertical line in Figure 2). Therefore, growth of Ge films did not lead to the formation of the Si–Ge bonds at the Ge/Si$_3$N$_4$ interface.

## 3.2. FTIR spectroscopy

IR absorbance spectra recorded in the range from 650 to 1350 cm$^{-1}$ for the samples with 200 nm Ge layers grown at temperatures from 30 to 500°C are shown in Figure 3(a). The spectra have been normalized to the maximum of the peak at 1090 cm$^{-1}$. They characterize Ge layers having amorphous (30°C) and polycrystalline (400 and 500°C) structure. All the spectra are seen to have similar shapes. It should be noticed that the IR absorbance spectra resemble ones we have previously observed for the Si films grown on the similar substrates in the similar conditions [7].

The spectra are characterized by two bands reaching maximums at ~825 and ~1090 cm$^{-1}$, which are associated with vibrations of the Si–O bonds. The stronger absorption band peaked at ~1090 cm$^{-1}$ has a short-wavelength wing, which extends up to ~1300 cm$^{-1}$, that is attributed to the vibrations of the Si–O and N–H bonds [23],[24].

Figure 3(b) presents an example of the spectra deconvolution using the Gaussian profiles of peaks. Each spectrum is seen to be constituted of eight spectral components peaked at about 735, 805, 860, 965, 1075, 1100, 1150 and 1190 cm$^{-1}$. The peaks at ~ 735, 805, 965, 1075, 1100 and 1190 cm$^{-1}$ are assigned to the vibrations of the Si–O bonds [25],[26]. The peak at ~ 1150 cm$^{-1}$ is generated by the vibrations of the N–H bonds [24]. The band around 860 cm$^{-1}$ relates to the absorption of radiation by the Si–N bonds [24],[27]. The comparison of the spectra has revealed that only absorption bands related to the vibration of the N–H (1150 cm$^{-1}$), Si–N (~ 860 cm$^{-1}$) and Si–O (~ 960 cm$^{-1}$) bonds had changed their intensities as a result of the germanium film deposition. Figure 3(c) presents growth temperature dependence of changes in the intensities of the Si–N, Si–O and N–H bands.

The growth of Ge films has led to a decrease in the intensity of the N–H absorption band compared to the spectra obtained at the initial Si$_3$N$_4$/SiO$_2$/Si(001) substrate. The intensity of the N–H absorption band has reduced with the growth of temperature. There is a noticeable difference between the intensity value of this band registered in the amorphous and polycrystalline Ge films. It is seen that the N–H absorption bands registered in the α-Ge samples grown at the temperatures of 30 and 120°C match exactly (numeral 1 in Figure 3(c)) but for the poly-Ge samples produced at the temperatures 400, 450, 500 and 600°C (numerals 2 and 3 in Figure 3(c)) these peaks coincide only roughly. The intensity of the N–H absorption band obtained for the initial $Si_3N_4/SiO_2/Si(001)$ substrate is seen to be the highest among those for all the samples. The similar result was earlier obtained by us for the samples with the Si films [7].

The change of the intensity of the Si–N absorption band at ~ 860 $cm^{-1}$ has occurred in a complex way. At first, for the samples grown at 30 and 120°C, its value has increased up to the maximum value (numeral 1 in Figure 3(c)). Then, with increasing growth temperature, the intensity has decreased and reached the minimum value at 500 and 600°C (numeral 3 in Figure 3(c)). The growth at 400 and 450°C has led to an intermediate value of the Si–N absorption band intensity (numeral 2 in Figure 3(c)). The intensity of the Si–N absorption band of the initial $Si_3N_4/SiO_2/Si(001)$ substrate has the least value among all the studied samples. Previously, in the samples with the grown Si films, we have observed gradual increase of the Si–N absorption band intensity with increasing growth temperature [7].

In our previous article [7], we associated the emergence of the absorption band peaked at ~ 960 $cm^{-1}$ with the fundamental antisymmetric stretching vibrations of the $Si_2O_6^{4-}$ silicate units ($Q^2$ according to the Engelhardt's Q-notation [28]) formed in the $SiO_2$ layer of the initial substrate since the Si film deposition did not lead to a significant variation of its intensity. In case of the Ge film growth, the intensity of this absorption band has decreased by several times. We have not observed any obvious dependence of the changes of the absorption intensity on the growth temperature (Figure 3(c)). The intensity of the absorption band at ~ 960 $cm^{-1}$ has demonstrated the highest value among all the studied samples in the initial $Si_3N_4/SiO_2/Si(001)$ substrate. Now, we could update the above assumption and suppose that the absorption band at ~ 960 $cm^{-1}$ relates to absorption by the vibrations of the Si–O bonds in $SiO_x$ or oxynitride ($Si_xO_yN_z$) layers lying on the surface of $Si_3N_4/SiO_2/Si(001)$ substrate. The similar Si–O absorption band was observed in the $SiO_x$ films in the wavenumber range from ~ 940 to 1080 $cm^{-1}$ and its position depended on the composition, shifting from ~ 940 cm$^{-1}$ for $x < 1$ to ~ 1080 cm$^{-1}$ in SiO$_2$ films [29],[30].

The absorption bands related to the Ge–O bonds, which usually emerge in the range from 830 to 870 cm$^{-1}$ in the GeO$_x$ layers [31], have not been observed in this study. We have not observed the absorption band peaked at ~ 976 cm$^{-1}$ related to the vibration of the Ge–O–Si bonds, which was reported in the work [32], either. Perhaps, it is connected with a low content of the Ge–O and Ge–O–Si bonds that was below the FTIR detection limit.

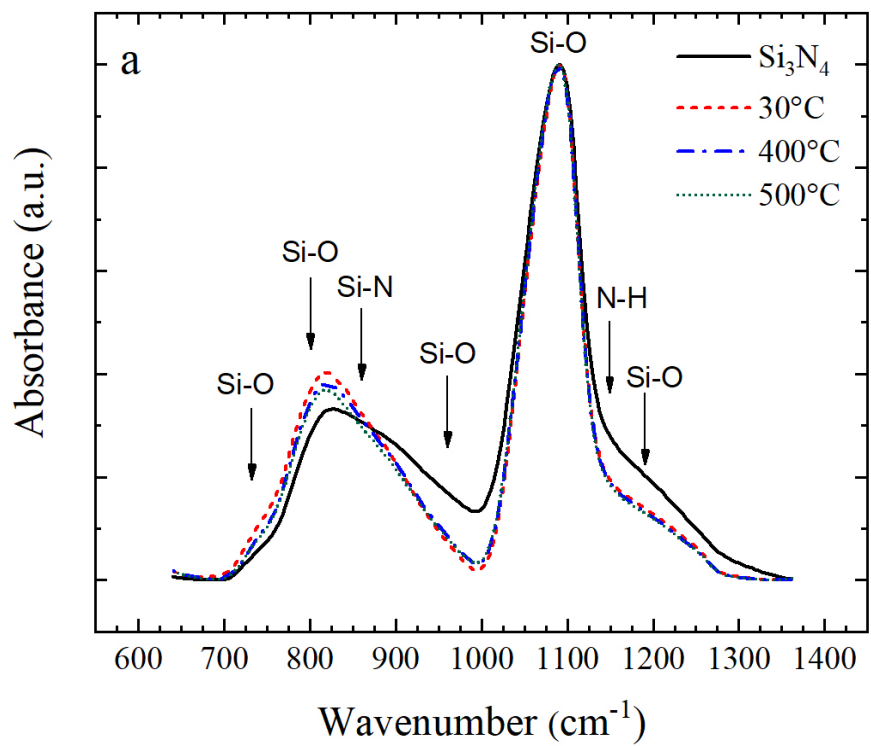

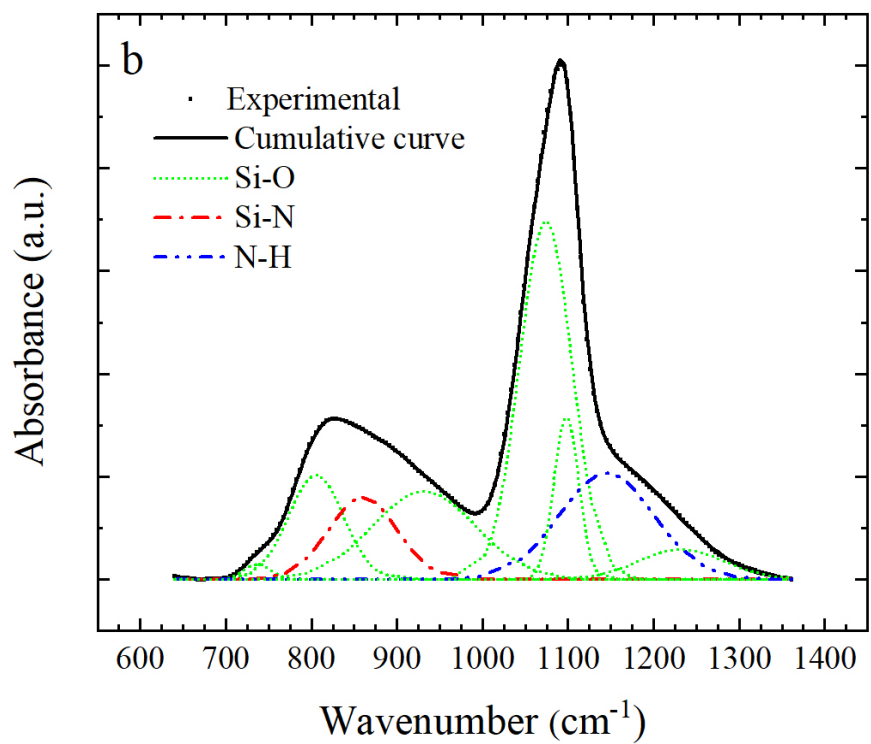

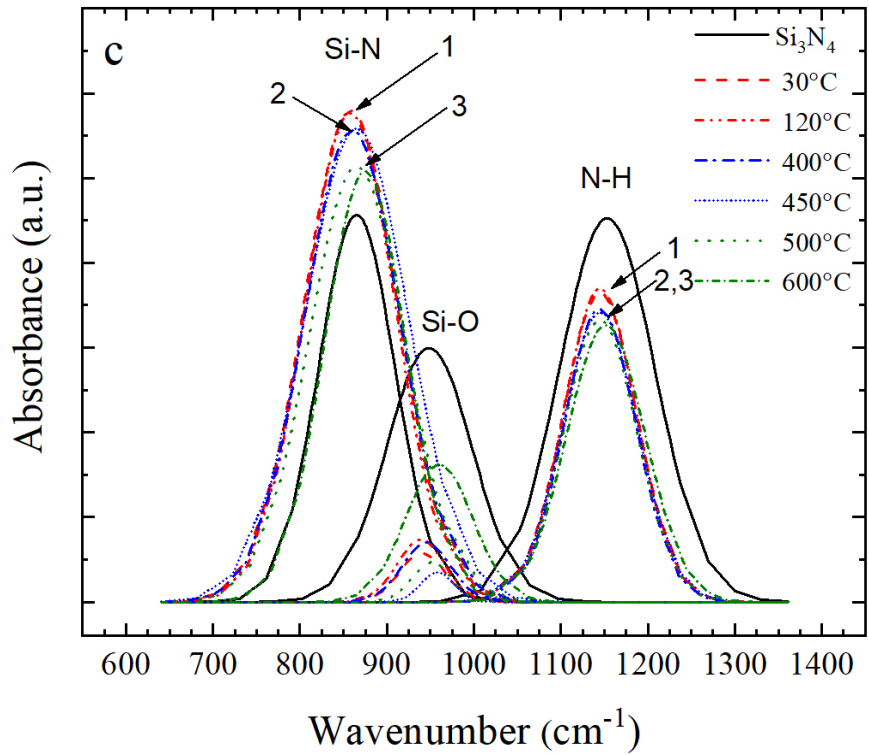

Figure 3. (a) IR absorbance spectra for the samples with 200 nm thick Ge films grown in the temperature range from 30 to 600°C; (b) an example of the deconvolution of spectrum shown in the panel (a) ($Si_3N_4$/$SiO_2$/Si(001) substrate) demonstrating eight spectral components related to the absorbance by the Si–O, Si–N and N–H bonds; (c) growth temperature dependence of changes in the intensities of Si–N, Si–O and N–H bands; the numerals correspond to the following growth temperatures: 30 and 120°C (1), 400 and 450°C (2), 500 and 600°C (3); the solid line marked as '$Si_3N_4$' corresponds to the spectrum for the $Si_3N_4$/$SiO_2$/Si(001) substrate.

Figure 4 presents the spectra of IR transmittance and reflectance sums for the samples with 200-nm thick Ge films grown in the temperature range from 30 to 600°C.

Remind that the transmittance and reflectance sum spectra enable the observation and the analysis of relatively weak absorption bands [7]. The spectra are seen to contain the absorption band at ~ 3300 cm$^{-1}$ (Figure 4) associated with the valence vibrations of the N–H bonds [24],[27]. The intensity of this peak is seen to decrease as the germanium deposition temperature increases. However, other absorption bands, which could appear within this wavenumber range, e.g., related to the vibrations of the Si–H ($\sim$ 2100 cm$^{-1}$) and Ge–H ($\sim$ 1880 см$^{-1}$) bonds [33],[34] have not been observed in this study (the usual positions of the latter bands in spectra are pointed out by the vertical lines in Figure 4). We should remark that we have previously observed the appearance of the line peaked at about 2100 cm$^{-1}$, connected with the Si–H bond vibration, owing to the silicon film deposition at low temperatures (30 and 160°C) [7].

The spectra demonstrated in Figure 4 have some peculiarities in the form of a deep decline of the sum spectra at wavenumbers greater than 2500 cm$^{-1}$. We assume that this is due to the impact of the light scatter by Ge film surface roughness on the reflection spectra since we have observed the transition from glossy to matte Ge film surface with the increase of the growth temperature, which could result in the partial loss of the reflected short-wavelength radiation. The contribution of light scattering in the short-wavelength region is large since the wavelength of light within it is close to the characteristic dimensions of the surface roughness (in the visible region of the spectrum, we have observed a matte surface of the structures). In the long-wavelength region, the light scattering loss decreases since the characteristic dimensions of the surface roughness become noticeably smaller than the wavelength.

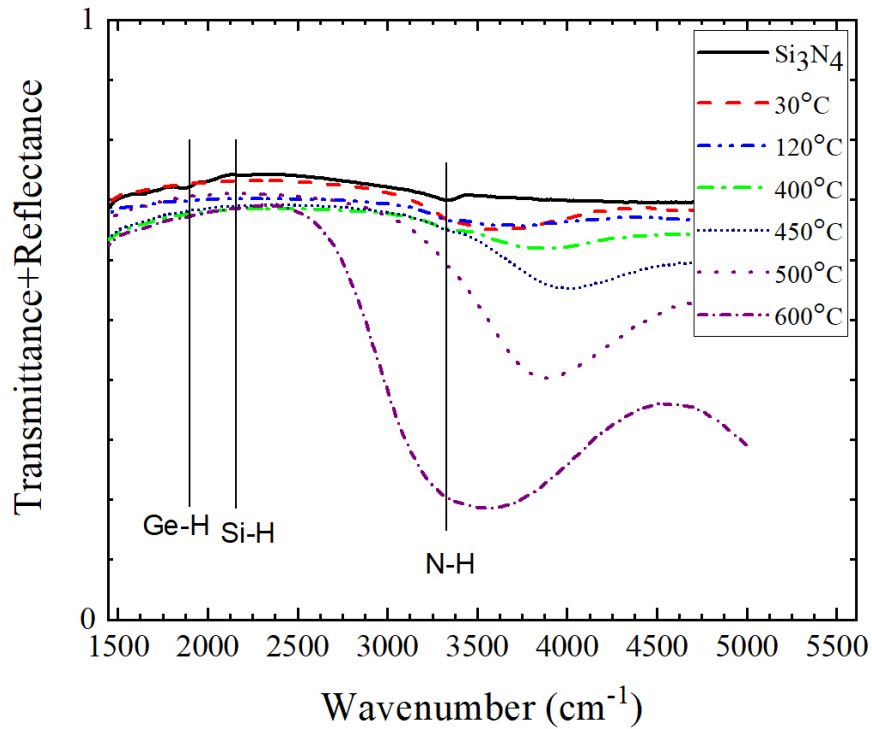

Figure 4. Spectra of IR transmittance and reflectance sums for the samples with 200 nm thick Ge films deposited in the temperature range from 30 to 600°C; the vertical lines point out the positions of the N–H, Si–H and Ge–H vibration bands; the solid line marked as 'Si$_3$N$_4$' corresponds to the spectrum for the Si$_3$N$_4$/SiO$_2$/Si(001) substrate.

The examination of the initial Si$_3$N$_4$/SiO$_2$/Si(001) substrates has shown that neither the preliminary chemical treatment in the ammonia-peroxide solution nor the preliminary annealing at 600°C in vacuum had introduced any modifications into FTIR spectra. Thus, one should conclude that the growth of the Ge films has been responsible for all the observed changes of FTIR spectra.

### 3.3. X-ray photoelectron spectroscopy

This section presents the study of X-ray photoelectron spectra of the initial Si$_3$N$_4$/SiO$_2$/Si(001) substrate and the sample coated with the 20-Å thin Ge layer grown at 500°C for 15 min. Figure 5 demonstrates a survey spectrum of the initial Si$_3$N$_4$/SiO$_2$/Si(001) substrate subjected to the preliminary chemical treatment in the ammonia-peroxide solution and the annealing at 600°C in vacuum. The spectrum is seen to contain peaks assigned to Si, N and O atoms.

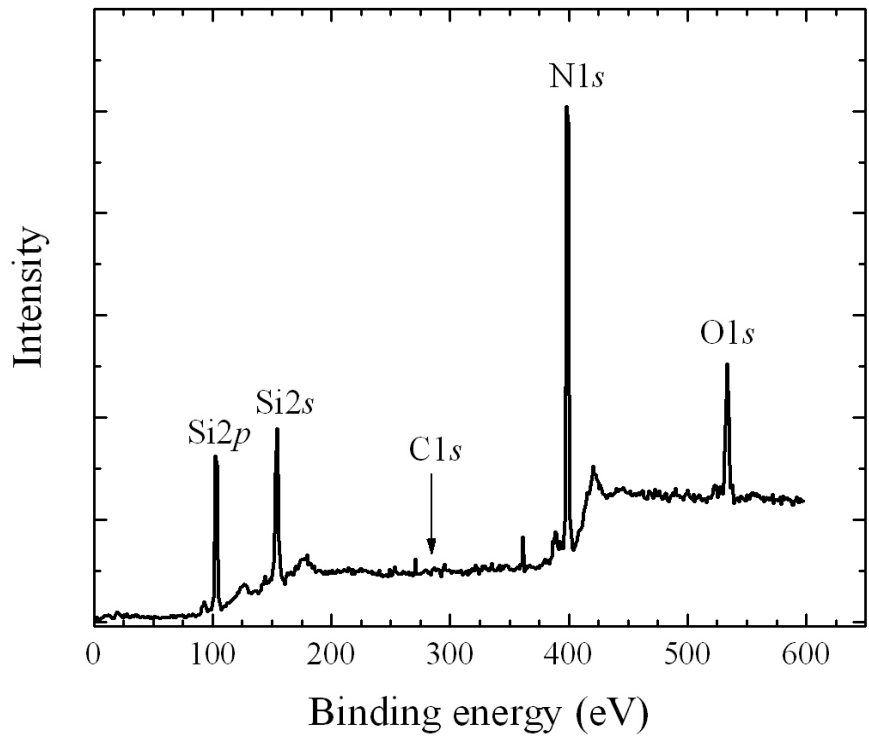

Figure 5. Survey X-ray photoelectron spectrum of the initial $Si_3N_4/SiO_2/Si(001)$ substrate.

Figure 6 demonstrates the results of the deconvolution of the $Si2p$ and $N1s$ peaks recorded from the surface of the initial $Si_3N_4/SiO_2/Si(001)$ substrate before and after $Ar^+$ ion etching. The $Si2p$ spectrum of the initial sample (Figure 6(a)) is split into two components with $E_b = 102.4$ eV (Si–N bonds in $Si_3N_4$) and $E_b = 103.5$ eV (Si–O bonds) related to Si–O–Si and/or O–Si–N bonding states of the oxynitride layer [35]-[37]. The $Si2p$ spectrum of the etched $Si_3N_4/SiO_2/Si(001)$ sample (Figure 6(b)) contains two components with $E_b = 102.2$ eV (Si–N bonds in $Si_3N_4$) and $E_b = 101.2$ eV ($Si^+$ bonding state). The peak with $E_b = 103.5$ eV (Si–O) disappeared as a result of etching.

The $N1s$ spectrum of the initial $Si_3N_4/SiO_2/Si(001)$ sample (Figure 6(c)) is subdivided into two components peaked at $E_b = 398.6$ eV associated with the N–Si bonds in $Si_3N_4$ and $E_b = 399.7$ eV presumably relating to the complex peak caused by vibrations of the N–O and N–H bonds (N–O/N–H peak) in the $Si_3N_4$ or oxynitride layers [38],[39]. Etching has not resulted in noticeable changes of the N1$s$ peak (Figure 6(d)), and it has kept on consisting of the same components located at $E_b$ = 398.6 and 399.7 eV but the peak area ratio of its components $S_{N-Si}/S_{N-O/N-H}$ has slightly increased. We suppose that the decrease in the N–O/N–H peak area ($E_b$ = 399.7 eV) may be elucidated by removing of its N–O component due to etching. Thus, we conclude that the native oxynitride layer has formed on the surface of the initial $Si_3N_4/SiO_2/Si(001)$ substrate because of the long-term storage in the air.

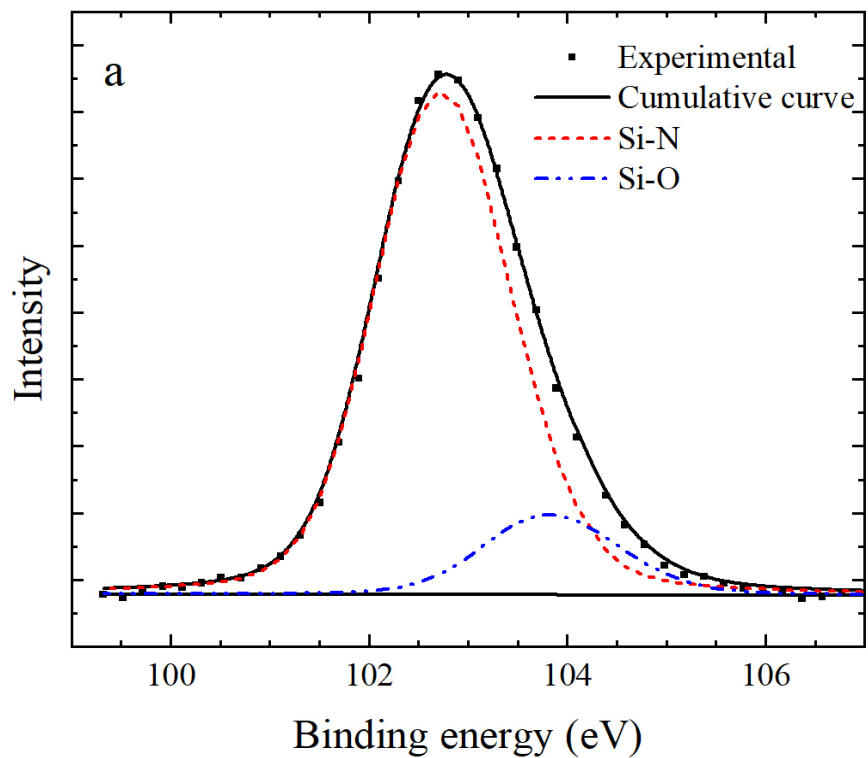

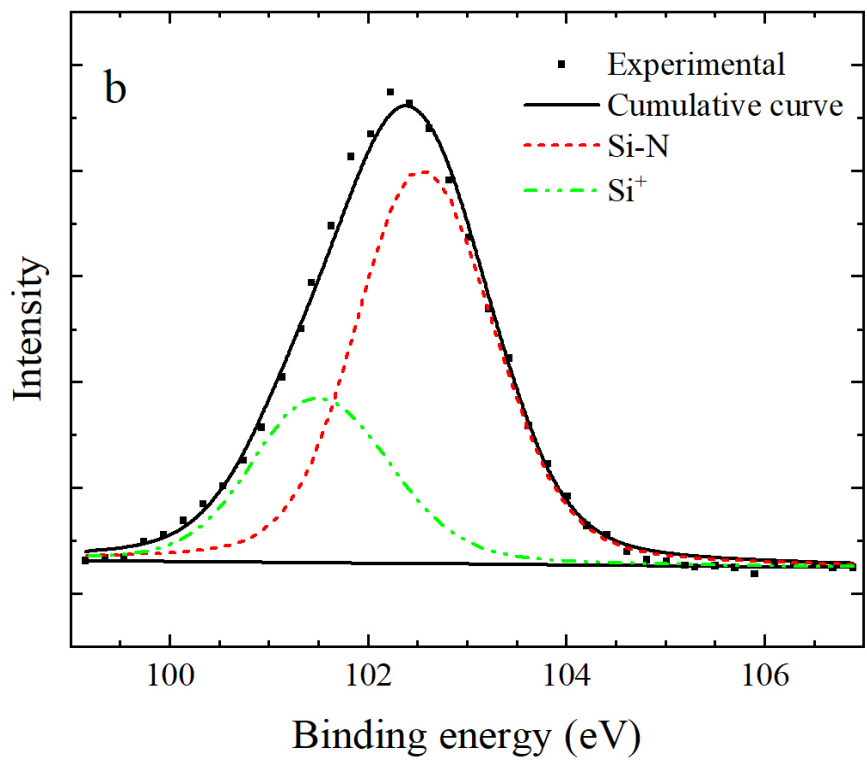

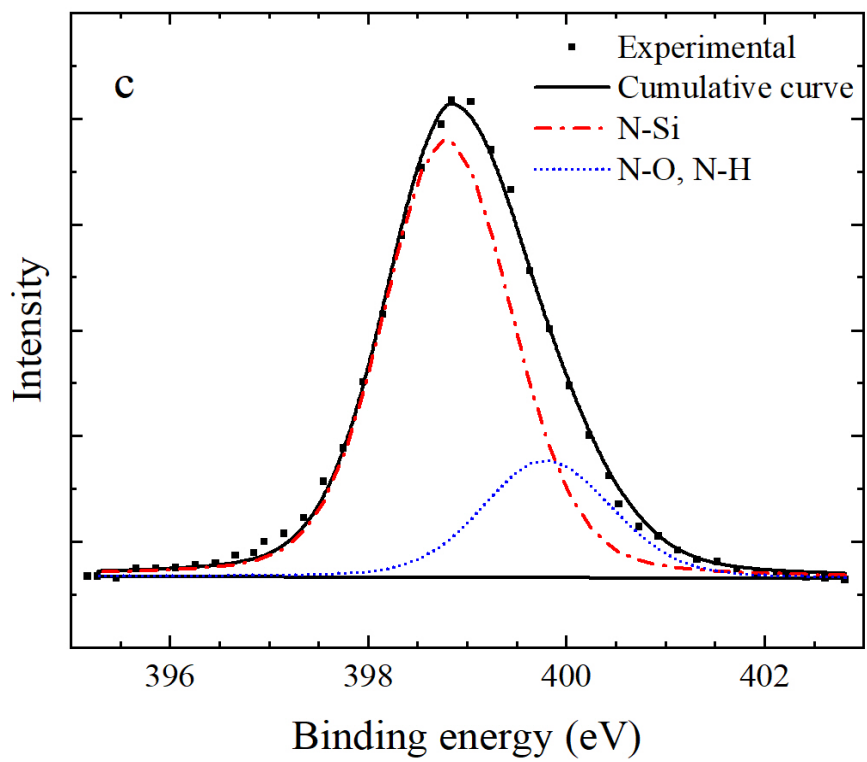

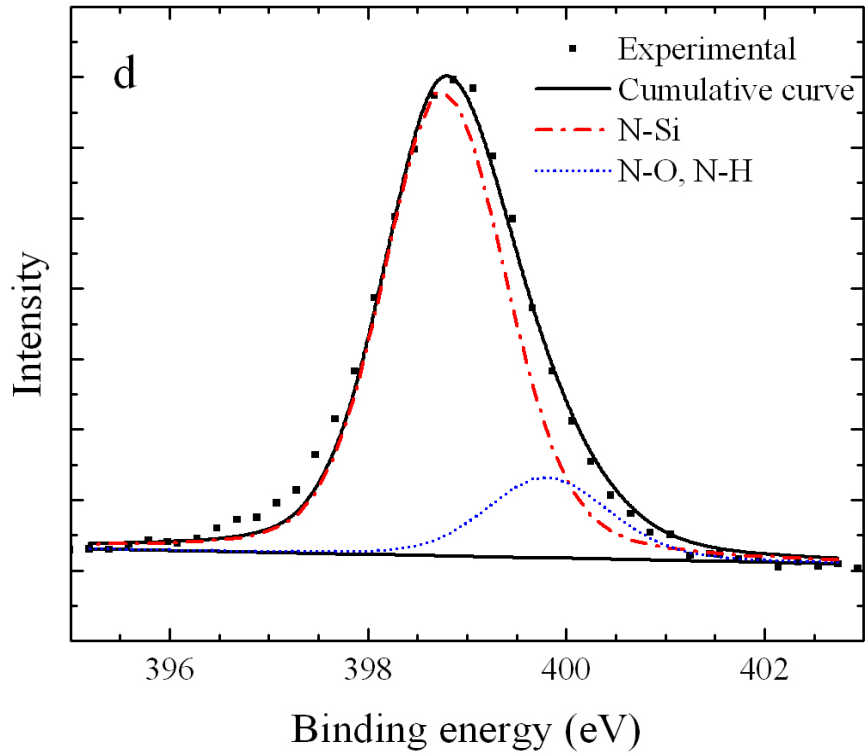

Figure 6. X-ray photoelectron spectra in the vicinity of the Si2$p$ (a,b) and N1$s$ (c,d) peaks of the Si$_3$N$_4$/SiO$_2$/Si(001) sample: (a) and (c) initial; (b) and (d) after Ar$^+$ ion etching.

Figure 7 demonstrates X-ray photoelectron spectra obtained at the sample of the Si$_3$N$_4$/SiO$_2$/Si(001) substrate coated by a thin (20 Å thick) Ge film grown at 500°C. The N1$s$ peak deconvolution shows that it is composed of three components with the binding energies $E_b$ = 397.5, 398.6 and 399.5 eV corresponding to the N–Ge, N–Si bonds and the N–O/N–H peak, respectively. The Si2$p$ peak consists of three components with the binding energies $E_b$ = 101.6, 102.6 and 103.6 eV associated with the Si$^+$, Si–N and Si–O bonding states, respectively. Note that the Si–Si (or Si–Ge) state with $E_b$ ~ 99 eV was not observed. The Ge3$d$ band is composed of two components peaked at $E_b$ = 29.7 and 30.4 eV related to the Ge–Ge and Ge–N bonds, respectively [40],[41]. The Ge2$p$ peak is decomposed into four components with the binding energies $E_b$ = 1218.1 eV, 1219.8 eV, which correspond to the Ge–Ge, Ge–N bonds and 1220.6 eV, and 1221.4 eV, which relates to Ge–O bonds [42],[43],[44]. The Ge2$p$ peak is characterized by lower electron attenuation length (EAL) than the Ge3$d$ one. According to computations presented in [45], EAL of the Ge2$p$ peak is ~ 8 Å and that of the Ge3$d$ one is ~ 28 Å. Thus, the Ge2$p$ peak is more sensitive to the composition of the near-surface domain of a sample and we have succeeded to detect a weak signal corresponding to Ge–O bonds in the spectrum. In the Ge3d peak, there is a great influence of the volumetric part of the sample, because of which we have failed to observe the Ge–O bonds of the near-surface region in this band. Therefore, we can suppose that the Ge–O and Ge–N bonds have been formed in the oxynitride layer due to the diffusion of Ge atoms. The feature of the Ge2$p$ and Ge3$d$ peaks is the predominance of the Ge–N components over all the others that corresponds to a higher content of the Ge–N bonds compared to the Ge–O ones. This enables supposing that Ge atoms have penetrated in the Si$_3$N$_4$ layer.

For the exploration of the composition along the sample depth, we made profiling by ion etching. To form a sample of a small size appropriate for etching, it was necessary to take it on air that was leading to the oxidation of its surface. The Ge–O bonds similar to those of bulk GeO$_2$ layers have been observed in both Ge2$p$ and Ge3$d$ peaks [42]–[44]. With the increase of the etching time, we have observed the decrease of the intensities of the Ge2$p$ and Ge3$d$ peaks up to the background level and it has been impossible to make peak analysis of these spectra. The N1$s$ and Si2$p$ peaks have slightly changed their intensities after etching and their components have modified as follows: the Si–O component has disappeared with etching in the Si2$p$ peak; in the N1$s$ peak, the component corresponding to the N–Ge peak has continued to be present after disappearing of the N–O/N–H peak and the Si–O one in the Si2$p$ peak but its intensity has decreased (Figure 8).

The appearance of the component attributed to the Si$^+$ in the Si2$p$ line has been observed in the initial samples as well as those coated with the Ge layer and etched by the bombardment of the Ar$^+$ ions (Figure 6b, Figure 7b and Figure 8b). The penetration in the subsurface area of the Ar$^+$ ions or the Ge atoms might lead to generation of the crystal lattice distortion and result in the transition of Si atoms to the new Si$^+$ bonding state. Note in support of the above assumption that the Si$^+$ bonding state has previously been shown to be formed as a result of processing of SiO$_{1.3}$ layers by Ar$^+$ ions [46].

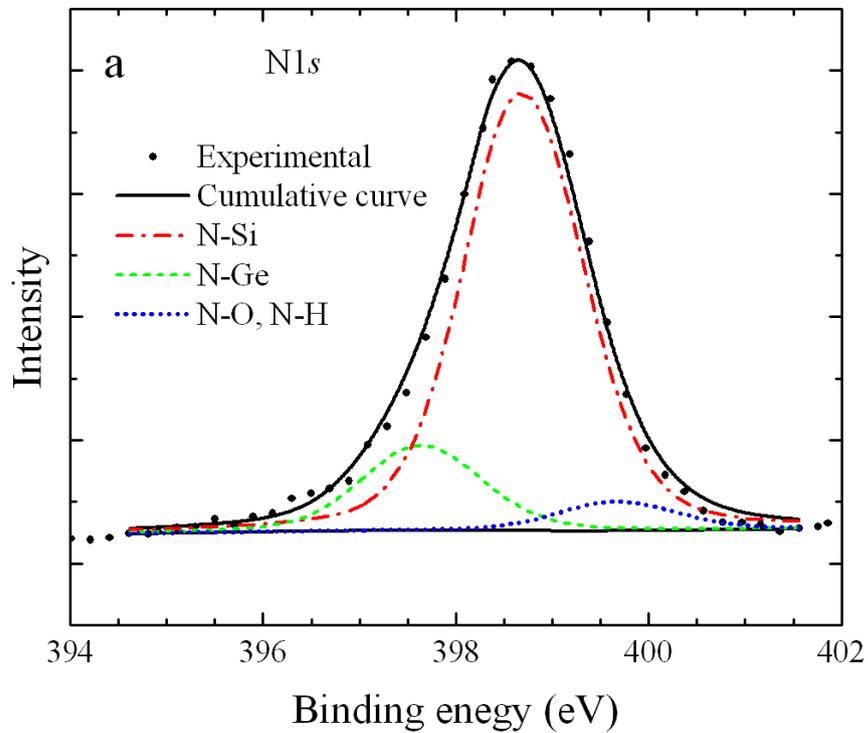

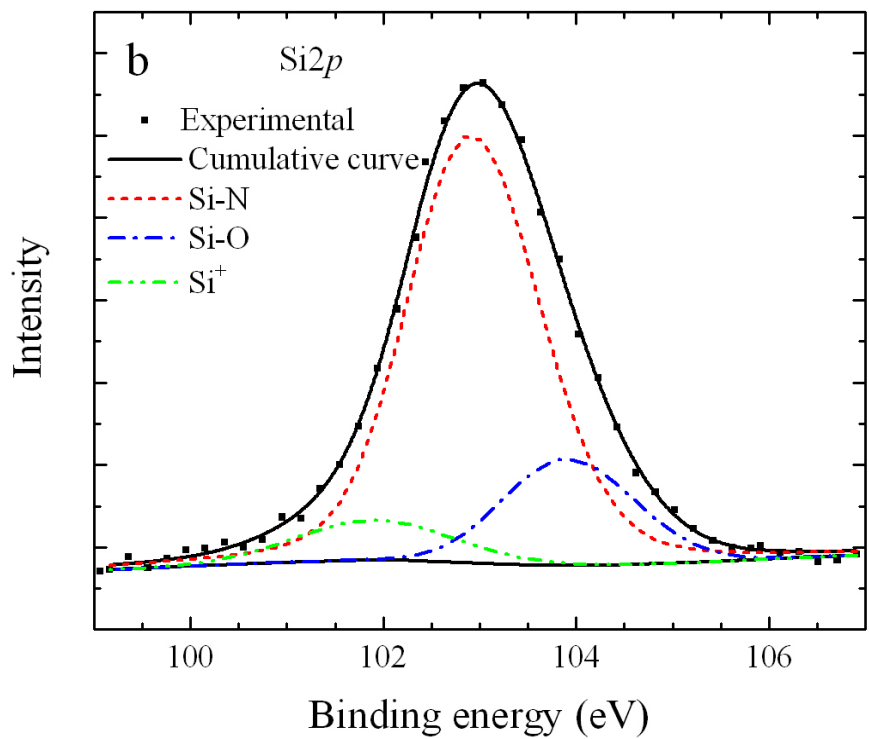

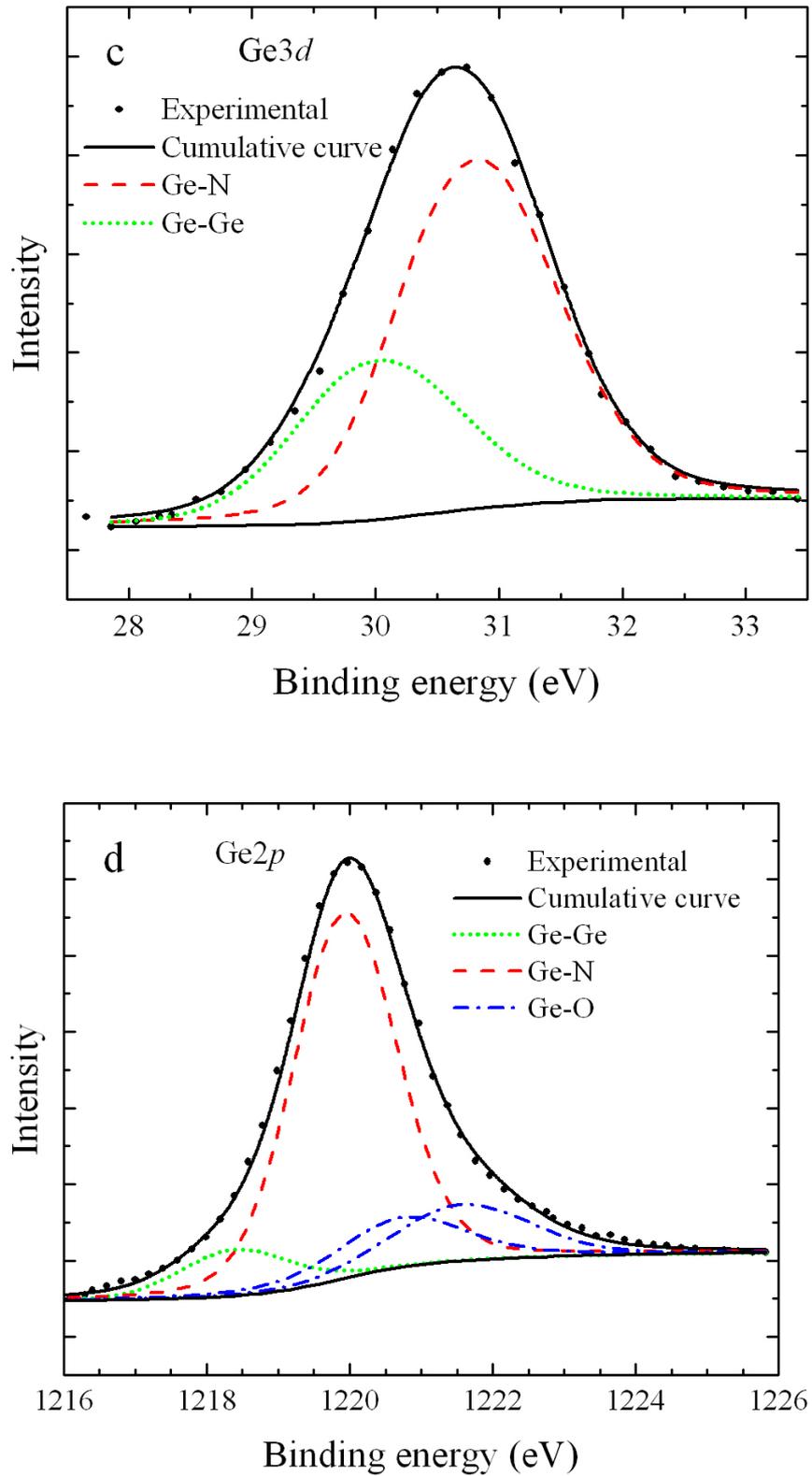

Figure 7. X-ray photoelectron spectra in the vicinity of the N1$s$ (a), Si2$p$ (b), Ge3$d$ (c) and Ge2$p$ (d) peaks obtained at the sample with 20-Å thick Ge layer grown at 500°C.

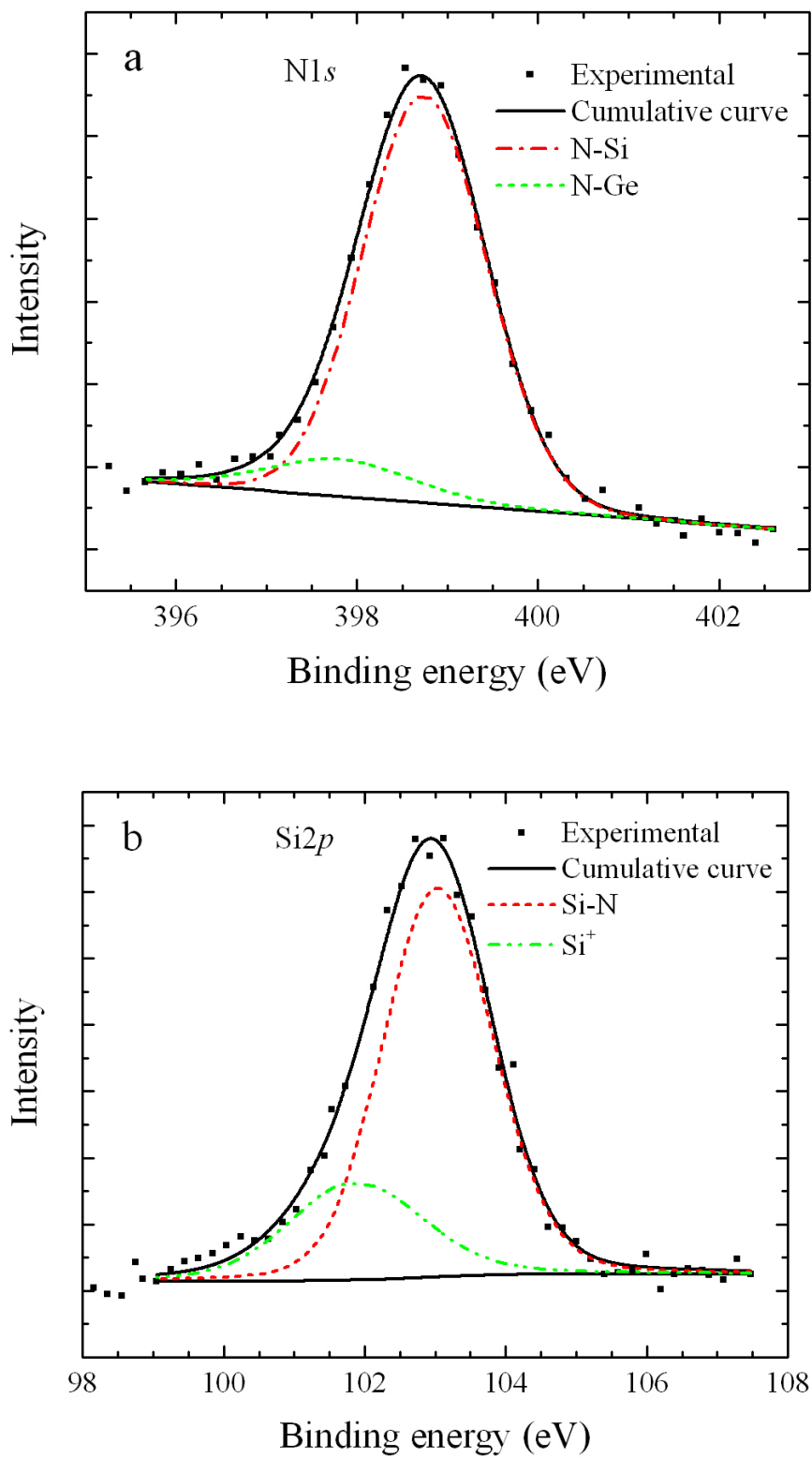

Figure 8. X-ray photoelectron spectra in the vicinity of the N1*s* (a) and Si2*p* (b) peaks obtained after etching of the sample with 20-Å thick Ge layer.

## 4. Discussion

The $Si_3N_4$ layers formed by CVD at high temperature are known to contain up to 8% atoms of hydrogen [12],[13]. According to our experimental results, the initial $Si_3N_4/SiO_2/Si(001)$ substrates contained the N–H bonds (Figs. Figure 3 and Figure 4) and the deposition of the Ge films as well as the Si ones [7] has led to the decrease of the intensity of the N–H absorption bands; at the same time, the absorption bands of the Ge–H and Si–H bonds are not observed in the spectra (Figure 4). We suppose that breaking of the N–H bonds in the $Si_3N_4$ layer and the diffusion of H atoms into the growing Ge film took place as it was previously observed for the growth of Si layers at the similar growth parameters [7]. The diffusion of hydrogen atoms into the growing Ge film has been observed already for the Ge deposition temperature of 30°C. Hydrogen atoms having passed in the Ge layer diffused to the growth surface and desorbed from it. The temperature of hydrogen desorption from Ge layers is known to be lower than from Si ones [47],[48] that could explain the absence of the Ge–H absorption bands in the FTIR spectra related to the samples grown even at rather low temperature such as 30°C.

To explain the above observation, we apply the model of the hydrogen diffusion proposed in Refs. [49]–[52] that was previously applied by us to the growth of the Si layers [7]. The diffusion of hydrogen atoms starts at low temperature, which cannot provide breaking such bonds as N–H and Si–H having bonding energy of ~ 4 and ~ 3 eV, respectively [53]. The activation energy of the diffusion of hydrogen atoms in α-Ge layers is ~ 1.5 eV [54] that slightly differs from the value in α-Si films [52]. The main mechanism of this diffusion into α-Ge films, similar to α-Si, is the movement of hydrogen atoms through interstitial states accompanied by the rupture of weak Ge–Ge bonds. A process limiting step is capturing of H atoms by dangling bonds of Ge atoms with the formation of Ge–H bonds (~ 3 eV [47], [54] ), which are deep traps.

Basic points of the model are a hydrogen atom chemical potential, the position of which depends on the concentration of hydrogen atoms in a layer, and a transport level characterizing interstitial positions of hydrogen atoms. A hydrogen atom passes from the level of deep trap to the transport level and can diffuse until it meets another deep trap. Since the level of $\mu_H$ depends on the content of hydrogen atoms, the latter can diffuse from the material with a higher level of $\mu_H$ into the material with a lower level. In the case in question, the $Si_3N_4$ layer has higher content of hydrogen atoms and higher level of $\mu_H$ than the growing film. Therefore, the hydrogen diffusion is directed from the $Si_3N_4$ layer into the growing Ge film resulting in the depletion of $Si_3N_4$ in hydrogen close to the $Ge/Si_3N_4$ interface. The thickness of the hydrogen depletion layer in $Si_3N_4$ appearing due to its outmigration expands as the deposition temperature increases.

Besides, a considerable mechanical stress might present in layered structures, which significantly weaken atomic bonds [55]. This may be considered as an essential mechanism that facilitates overcoming of Ge–Ge (Si–Si) and Ge–H (Si–H and N–H) bond breaking barrier [7] and hence, the hydrogen diffusion. Notice, that a mechanical stress as a driving force of the process intensifies the diffusion at a rather low film growth temperature just like it promotes fracturing or depolymerization of glass controlling its long-term corrosion at room temperature [56],[57].

It should be noticed also that the hydrogen diffusion should occur in different ways during the growth of amorphous and polycrystalline films; a film thickness should also have an effect on this process. These assumptions need further studies, however.

The results obtained by the X-ray photoelectron spectroscopy have revealed the formation of the Ge–N bonds in the samples with the Ge layers (Figure 6). Based on the FTIR and XPS results, we suppose the Ge–N bonds to be formed in the oxynitride and $Si_3N_4$ layers as a result of the diffusion of Ge atoms from the growing film. The following results support this assumption. First, the Si–O absorption band ($\sim 960$ cm$^{-1}$) in IR absorption spectra are associated with the natural oxynitride layer, which formed during the storage of the initial $Si_3N_4/SiO_2/Si(001)$ substrate in air for a long time [58],[59]. This layer was removed by $Ar^+$ ion etching (Figure 6a and b) but the Ge deposition did not lead to its decomposition (Figure 7b). Simultaneously, $Si^+$ bonding state connected with raising disorder in the surface layer appeared in the XPS spectrum of Si2$p$. Second, in Figure 9, we have presented the IR absorption spectra of the samples with 200-nm thick poly-Si [7] and poly-Ge films deposited at 500°C. The comparison of the spectra shows that the growth of the Ge layer has significantly reduced the strength of the Si–O absorption band (~960 $cm^{-1}$) yet the growth of the Si layers has changed it much less. Thus, the growth of the Si and Ge films influences the oxynitride layer differently. We assume that Ge atoms diffused into the $Si_3N_4$ layer, passed through the oxynitride layer and deformed it that resulted in the loss of the Si–O bonds IR activity. Si atoms having a smaller size did not produce so considerable effect. Third, according to the X-ray photoelectron spectra of the Ge2$p$ peak (Figure 7d), the Ge–O bonds formed in the near-surface region, and the Ge–N bonds contributed the most significantly to the Ge2$p$ and Ge3$d$ peaks. Fourth, the depth profiling of the sample with 20-Å thick poly-Ge layer has shown that the Si–O component has disappeared from the Si2$p$ X-ray photoelectron spectra whereas the Ge–N component has remained in the N1$s$ one (Figure 8). Thus, the Ge–N bonds could form under the natural oxynitride layer. Fifth, let us consider the changes in intensities of the Si–N absorption bands (Figure 9). In the sample with the Si film, its intensity is seen to increase more noticeably than in the one with the Ge film. The comparison of the growth temperature dependence of the intensity of the Si–N absorption band reveals that it decreased with increasing growth temperature of the Ge layers (Figure 3) but increased during the growth of Si films [7]. We suppose that such behavior could also be explained by the diffusion of Ge and Si atoms from the growing films into the $Si_3N_4$ layer. The raise of the Ge (Si) film growth temperature leads to the increase in numbers of hydrogen atoms leaving the $Si_3N_4$ layer and intensifies the diffusion processes. After breaking of the N–H bonds and the diffusion of the hydrogen atom, there appear nitrogen atoms having a dangling bond, which could be closed by forming a new Si–N bond between the nitrogen atom and the nearest Si atom. During the deposition of the Si film, Si atoms diffuse into the $Si_3N_4$ layer and form new Si–N bonds; as a result, the intensity of the Si–N absorption band increases. During the deposition of the Ge film, Ge atoms diffuse into the $Si_3N_4$ layer and form Ge–N bonds; as a result, the intensity of the Si–N absorption band decreases. With the increase of the growth temperature, the formation of new bonds between the nitrogen atoms having dangling bonds and the diffusing Ge (Si) atoms prevails over the bonding to the nearest Si atoms. The important question is what diffusion way of the Ge (Si) atoms is. We suppose they move on the places that have become vacant after the escape of hydrogen atoms. This is similar to the substitution or the diffusion by vacancies of H atoms since bigger Ge atoms cannot easily diffuse interstitially in $Si_3N_4$. Thus, the diffusion of Ge or Si atoms takes place only simultaneously with the diffusion of hydrogen atoms from the $Si_3N_4$ layer. A possible driving factor of this process is the formation of the Ge–N (~ 2.7 eV) and Si–N (~ 3.1 eV) bonds, which possess a higher bonding energy than the Ge–Ge (~ 1.9 eV), Si–Si (~ 2.3 eV) or Ge–Si ones [60]. A peculiarity of the presented FTIR spectra is the absence of the absorption bands corresponding to the vibrations of the Ge–N bonds, which were observed for $GeN_x$ layers in the range from ~ 690 to 750 см$^{-1}$ [60]–[62] depending on their composition. This might be caused by low number density of the Ge–N bonds or by loosing of IR activity since Ge atoms forming the Ge–N bonds could introduce mechanical stresses that distorted the lattice resulting in a high dispersion of the bond length and angle between them.

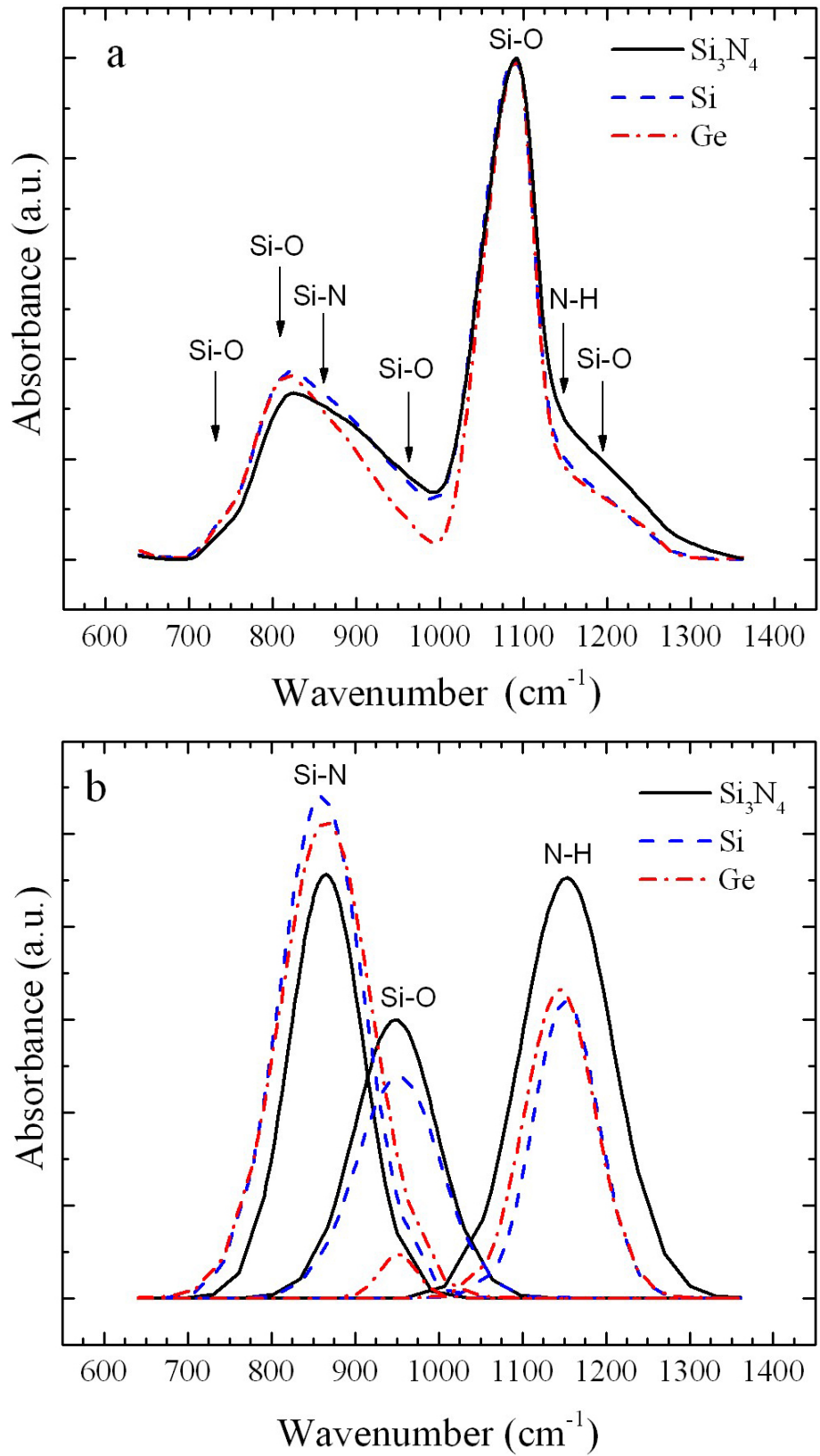

Figure 9. (a) IR absorbance spectra for the samples with 200-nm thick Ge and Si films grown at the temperature of 500°C; (b) the results of the deconvolution of the spectra shown in the panel (a); the solid line marked as 'Si₃N₄' represents the spectrum for the Si₃N₄/SiO₂/Si(001) substrate.

## 5. Conclusion

The deposition of the Ge films on the dielectric $Si_3N_4$ substrates containing hydrogen atoms has led to the diffusion of hydrogen atoms from the dielectric layer into the growing film and desorbed from its surface. This process took place within the temperature range from 30 to 600°C and cannot be explained as a thermally activated process only. We have observed that the intensity of the IR absorption bands related to the vibrations of N–H and Si–N bonds reduced with increase of growth temperature. Simultaneously, the peaks assigned to the Ge–N and Ge–O bonds appeared in the X-ray photoelectron spectra. We suppose that it was induced by the diffusion of the Ge atoms into the natural oxynitride layer covering the surface of the initial $Si_3N_4/SiO_2/Si(001)$ substrate and the dielectric $Si_3N_4$ one. The peak area ratios $S_{Ge-N}/S_{Ge-O}$ and $S_{Ge-N}/S_{Ge-Ge}$ of components in the Ge2$p$ and Ge3$d$ peaks allow us to suppose that Ge atoms could diffuse into the $Si_3N_4$ layer passing though oxynitride and forming the Ge–N bonds there. In case of the deposition of the Si films [7], it could be expected that the similar processes took place and the Si atoms from the growing film diffused into the dielectric $Si_3N_4$ layer. This led to the increase in the intensity of the IR absorption bands related to the vibrations of the Si–N bonds.

The processes taking place during deposition of the germanium (silicon) films on the $Si_3N_4$ dielectric layers can be elucidated within the diffusion model proposed in Ref. [49], which considered the difference in chemical potentials of hydrogen atoms in the dielectric layer and the germanium (silicon) film as the driving force of the diffusion. The process of the migration of hydrogen atoms into the growing germanium (silicon) film initiated the diffusion of Ge (Si) atoms in the opposite direction into the dielectric layer.

**Acknowledgments**

This research did not receive any specific grant from funding agencies in public, commercial or not-for-profit sectors.

The Center for Collective Use of Scientific Equipment of GPI RAS supported this research via presenting admittance to its equipment.

**Conflicts of interest.**

No conflicts of interest exist, which could potentially influence the work.